\documentclass[a4paper,11pt]{article}
\usepackage{pos}
\title{Multiscale evolution of heavy flavor in the QGP}

\author*[a]{Gojko Vujanovic}
\onbehalf{for the JETSCAPE Collaboration}

\affiliation[a]{Department of Physics, University of Regina, Regina, SK S4S 0A2, Canada}

\emailAdd{gojko.vujanovic@uregina.ca}

\abstract{Shower development dynamics for a jet traveling through the quark-gluon plasma (QGP) is a multiscale process, where the heavy flavor mass is an important scale. During the high virtuality portion of the jet evolution in the QGP, emission of gluons from a heavy flavor is modified owing to heavy quark mass. Medium-induced radiation of heavy flavor is sensitive to microscopic processes (e.g. diffusion), whose virtuality dependence is phenomenologically explored in this study. In the lower virtuality part of shower evolution, i.e. when the mass is comparable to the virtuality of the parton, scattering and radiation processes of heavy quarks differ from light quarks. The effects of these mechanisms on shower development in heavy flavor tagged showers in the QGP is explored here. Furthermore, this multiscale study examines dynamical pair production of heavy flavor (via virtual gluon splittings) and their subsequent evolution in the QGP, which is not possible otherwise. A realistic event-by-event simulation is performed using the JETSCAPE framework. Energy-momentum exchange with the medium proceeds using a weak coupling recoil approach. Using leading hadron and open heavy flavor observables, differences in heavy versus light quark energy-loss mechanisms are explored, while the importance of heavy flavor pair production is highlighted along with future directions to study.}

\FullConference{
 26-31 March 2023 \\
 Aschaffenburg, Germany\\}

\begin{document}
\maketitle
\section{Introduction}
JETSCAPE (Jet Energy-loss Tomography with a Statistically and Computationally Advanced Program Envelope) framework is a flexible publicly released event-generator framework providing theoretical modeling of all aspects of heavy ion collisions. While in this contribution focus is given towards the quenching of high energy partons, modules are present to cover other aspects of aspect relativistic heavy-ions, notably relativistic dissipative hydrodynamics and hadronic transport. Each module can be user-modified, giving any user of the JETSCAPE framework the ability to focus on a subset of physics simulations of interest, knowing that other aspects of the simulation are state of the art. Indeed, the JETSCAPE framework also supplies Bayesian analysis tools needed to accurately constrain physical processes. These Bayesian tools are actively being used \cite{JETSCAPE:2020mzn,JETSCAPE:2020shq,JETSCAPE:2021ehl,Chen:2023HP} to improve our understanding of relativistic heavy-ion physics. The work herein investigates how different physical mechanisms involved in jet quenching affect the nuclear modification factor $R_{AA}$. Specifically, this contribution explores (i) how dynamical generation of heavy quarks inside the QGP effects light and heavy flavor $R_{AA}$ and (ii) investigates the effects of a virtuality-dependent $\hat{q}$ on light and heavy flavor $R_{AA}$.

\section{Numerical simulations}
The simulation of parton interaction with the QGP uses a two-step approach for both light and heavy flavor partons: the high virtuality (and high energy) portion of parton quenching in the QGP is described using the higher twist formalism \cite{Wang:2001ifa,Majumder:2009ge} implemented in MATTER (Modular All Twist Transverse-scattering Elastic-drag and Radiation) \cite{Abir:2015hta,Majumder:2013re}, while the low virtuality (and high energy) portion of quenched partons is modeled via Linear Boltzmann Transport (LBT) \cite{Luo:2018pto}. 

MATTER contains both vacuum showering of highly virtual partons as well as in-medium corrections to the splitting function, which includes novel coherence effects giving a virtuality-dependent $\hat{q}$ first proposed in Ref.~\cite{Kumar:2019uvu}. The medium-modified splitting function of the process $Q\to Q+g$ --- which also holds for light flavor in the limit $M\to 0$ --- is \cite{Abir:2015hta}:
\begin{eqnarray}
\frac{dN^{\rm vac}}{dz dt} + \frac{dN^{\rm med}}{dz dt } &=& \frac{\alpha_s(t)}{2\pi} \frac{P_{g\gets Q}(z)}{t}\left\{ 1+ \int^{\tau^+_Q}_0 d\tau^+\frac{1}{z(1-z)t(1+\chi)^2} \left[2-2\cos\left( \frac{\tau^+}{\tau^+_Q} \right)\right]\right.\times\nonumber\\
&\times& \left. \left[  \left(\frac{1+z}{2}\right) - \chi + \left(\frac{1+z}{2}\right) \chi^2 \right]\hat{q}(t)  \right\}
.\nonumber\\
\label{eq:hq_ht}
\end{eqnarray} 
where $z$ is the momentum fraction of the daughter heavy quark, $M$ is the mass of the heavy quark, $\chi=(1-z)^2M^2/l^2_\perp$, with $l^2_\perp$ being the relative transverse momentum square between the outgoing daughter partons, determined via $z(1-z)t=l^2_\perp(1+\chi)$, while $t$ is the virtuality of the heavy quark and $P_{g\gets Q}(z)=C_F\left(\frac{1+z^2}{1-z^{\,\,\,}}\right)$ is the splitting function. The integral over light-cone time $\tau^+$ in Eq. (\ref{eq:hq_ht}) assumes the medium is in its rest frame, with the upper bound $\tau^+_Q=2q^{+}/t$ being given by the ratio of forward light-cone momentum $q^+=\left(q^0+{\bf q}\cdot \hat{n}\right)/\sqrt{2}$ (with $\hat{n}={\bf q}/\vert {\bf q}\vert$), and the virtuality $t$. The virtuality-dependent $\hat{q}(t)$ is parametrized as
\begin{eqnarray}
\hat{q}(t) &=& \hat{q}^{HTL}\frac{c_0}{1+c_1\ln^2(t)+c_2\ln^4(t)}\nonumber\\
\label{eq:qhat_t}
\end{eqnarray}
where $c_1$ and $c_2$ are tunable parameters, $t=E^2-{\bf p}^2-m^2$ is the virtuality of the parton with $p^\mu=(E,{\bf p})$, and $c_0=1+c_1\ln^2(t_s)+c_2\ln^4(t_s)$ is an overall normalization ensuring that the $t$-dependent contribution is unitless and lies within 0 and 1 as $t$ does not go below $t_s$, while the expression for $\hat{q}^{HTL}$ is in Ref.~\cite{JETSCAPE:2022hcb}. 

In addition to medium-modified radiation, MATTER includes $2\to2$ scatterings at leading order in the QCD coupling $\alpha_s$. An important process that can only be studied in a multistage model, such as MATTER+LBT, is the dynamical in-medium heavy flavor production. Thus, heavy flavor pair production via $g\to Q+\bar{Q}$, where $(\bar{Q})Q$ is the heavy (anti)quark, is the key process studied here. Indeed, a multiscale study, such as MATTER+LBT, is required to fully estimate the importance of dynamical in-medium heavy quark production. Of course, charm pair production is the process with the largest allowed phase space, and hence D-meson $R_{AA}$, in addition to charged hadron $R_{AA}$, will be explored below. Unlike recent calculations \cite{Attems:2022otp,Attems:2022ubu}, the present simulation relies on light-flavor $\hat{q}$ effects when correcting the $P_{Q\leftarrow g}$ splitting function. However, the medium-modified $P_{Q\leftarrow g}$ splitting function is taking into account kinematic cuts associated with heavy quark masses \cite{JETSCAPE:2022hcb}. This approximation for $P_{Q\leftarrow g}$ will be revisited in the future, as discussed below. In the low virtuality sector, the LBT simulation is used which also includes $2\to2$ scatterings, and employs the radiation kernel of MATTER within a $2\to3$ process to describe in-medium emissions (see Ref.~\cite{JETSCAPE:2022hcb} for details). As far as simulations of the QGP and its subsequent hadronic evolution is concerned, the Bayesian maximum {\it a posteriori} parameters are used \cite{Bernhard:2019bmu} to generate the QGP medium for Pb-Pb collisions at $\sqrt{s_{NN}}= 5.02$ TeV in the 0-10\% centrality class (cf. details in \cite{JETSCAPE:2022hcb}).

The observable of interest herein is $R_{AA}$:
\begin{eqnarray}
R_{AA}= \frac{\frac{d \sigma_{AA}}{d p_T}}{\frac{d \sigma_{pp}}{d p_T}}=\frac{\sum_{\ell} \frac{d N_{AA,\ell}}{d p_T}\hat{\sigma}_{\ell}}{\sum_{\ell} \frac{d N_{pp,\ell}}{d p_T}\hat{\sigma}_{\ell}}
\label{eq:R_AA}
\end{eqnarray}
where $\frac{dN_{AA}}{dp_T}$ and $\frac{dN_{pp}}{dp_T}$ are the multiplicity of either D-mesons or charged hadrons originating from A-A and p-p collisions in the experimentally given $p_T$ bin, respectively. The cross-section for producing the hard scattering process of the given range $\ell$ in transverse momentum $\hat{p}_T$ is $\hat{\sigma}_{\ell}$ ($\hat{p}_T$ is the transverse momentum of the exchanged parton in the hard scattering sampled by PYTHIA). Many $\hat{\sigma}_{\ell}$ are sampled to span the large kinematic range of the collision \cite{JETSCAPE:2022hcb}.

\section{Results}
\begin{figure}
\begin{center}
\begin{tabular}{cc}
\includegraphics[width=0.495\textwidth]{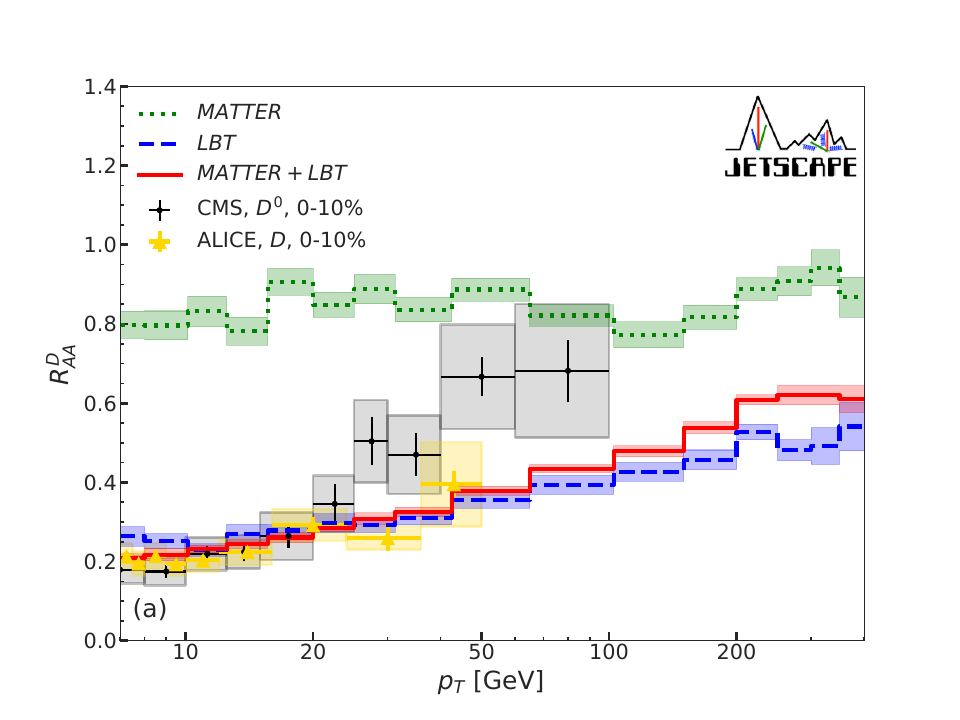} & \includegraphics[width=0.495\textwidth]{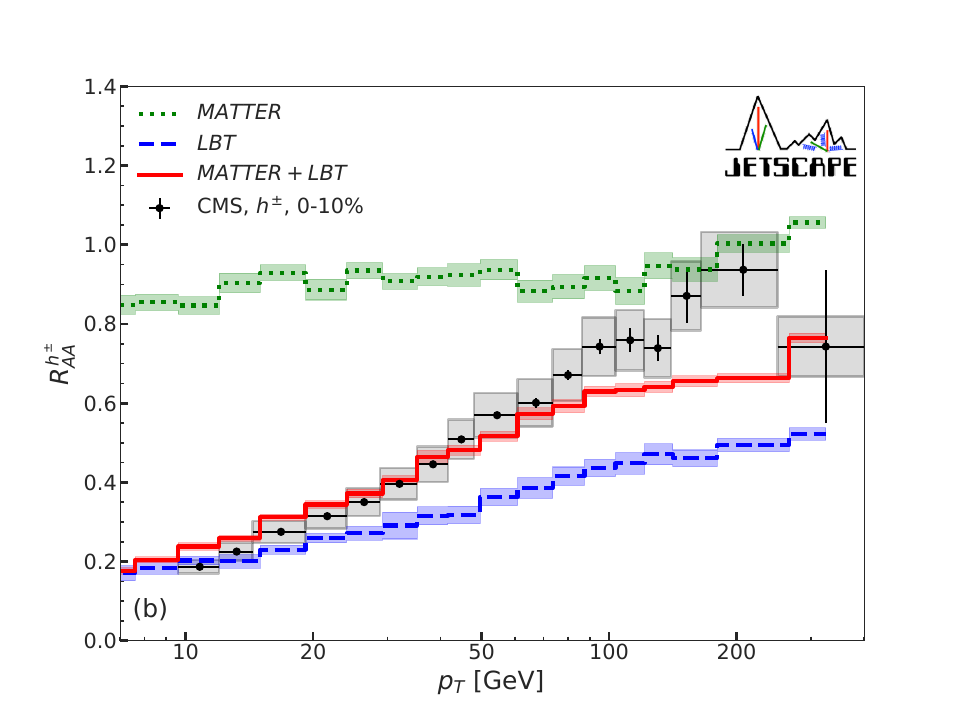}
\end{tabular}
\end{center}
\caption{Nuclear modification factor for D-mesons (a) and charged hadrons (b) at the $\sqrt{s_{NN}}=5.02$ TeV Pb-Pb collisions at the LHC in the 0-10\% centrality. $c_1=10,\ c_2=100$ are set within the $\hat{q}(t)$ parametrization [see in Eq.~(\ref{eq:qhat_t})] for the MATTER alone and the MATTER+LBT curve. The other parameter is the switching virtuality $t_s=4$ GeV$^2$ between MATTER and LBT, whose value is found to best describe the $R_{AA}$ data. Finally, the p-p baseline for the LBT curve is calculated using PYTHIA whereas the p-p baseline for the MATTER and MATTER+LBT cases are calculated using MATTER vacuum \cite{JETSCAPE:2019udz}. Data taken from Ref.~\cite{CMS:2017qjw,CMS:2016xef,ALICE:2018lyv}.}
\label{fig:MAT_vs_LBT_all}
\end{figure}
\begin{figure}
\begin{center}
\begin{tabular}{cc}
\includegraphics[width=0.495\textwidth]{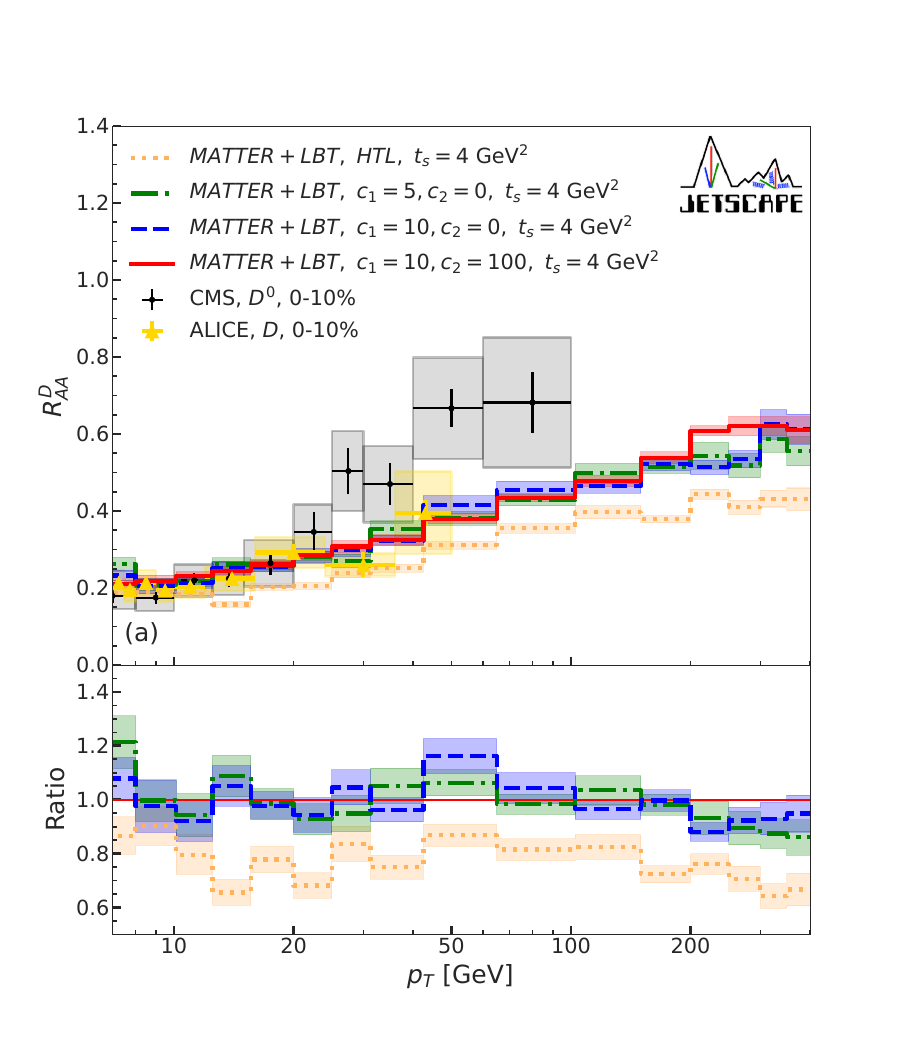} & \includegraphics[width=0.495\textwidth]{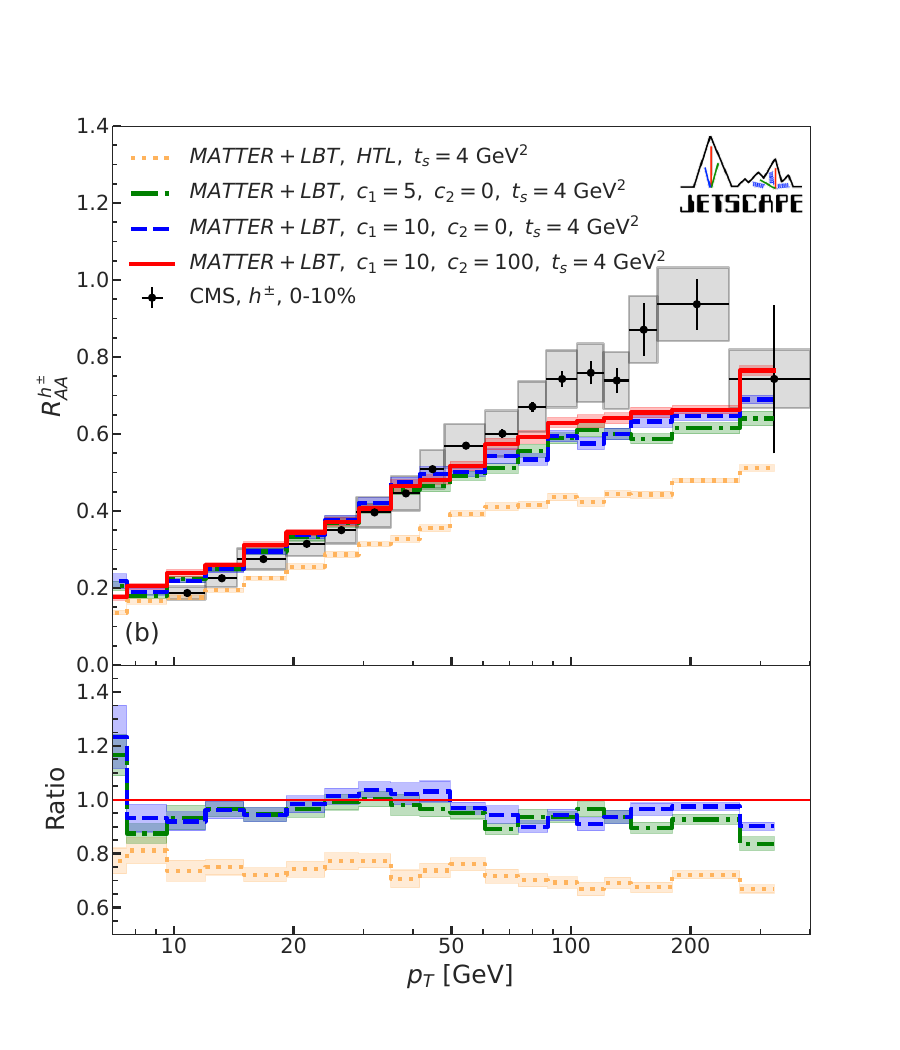}
\end{tabular}
\end{center}
\caption{(Color online) Nuclear modification factor for D-mesons (a) and charged hadrons (b) in $\sqrt{s_{NN}}=5.02$ TeV Pb-Pb collisions at the LHC at 0-10\% centrality. The parametrization of $\hat{q}(t)$ is varies with $c_1$ and $c_2$ values indicated. The ratio in the bottom plots are taken with respect to the $c_1=10, c_2=100$ case with $\hat{q}(t)$ parametrization [see Eq.~(\ref{eq:qhat_t})]. The sensitivity of $R_{AA}$ to $t_s$ is explored in Ref.~\cite{JETSCAPE:2022hcb}.}
\label{fig:MATTER_LBT_comp_qhat}
\end{figure}
\begin{figure}[!h]
\begin{center}
\begin{tabular}{cc}
\includegraphics[width=0.495\textwidth]{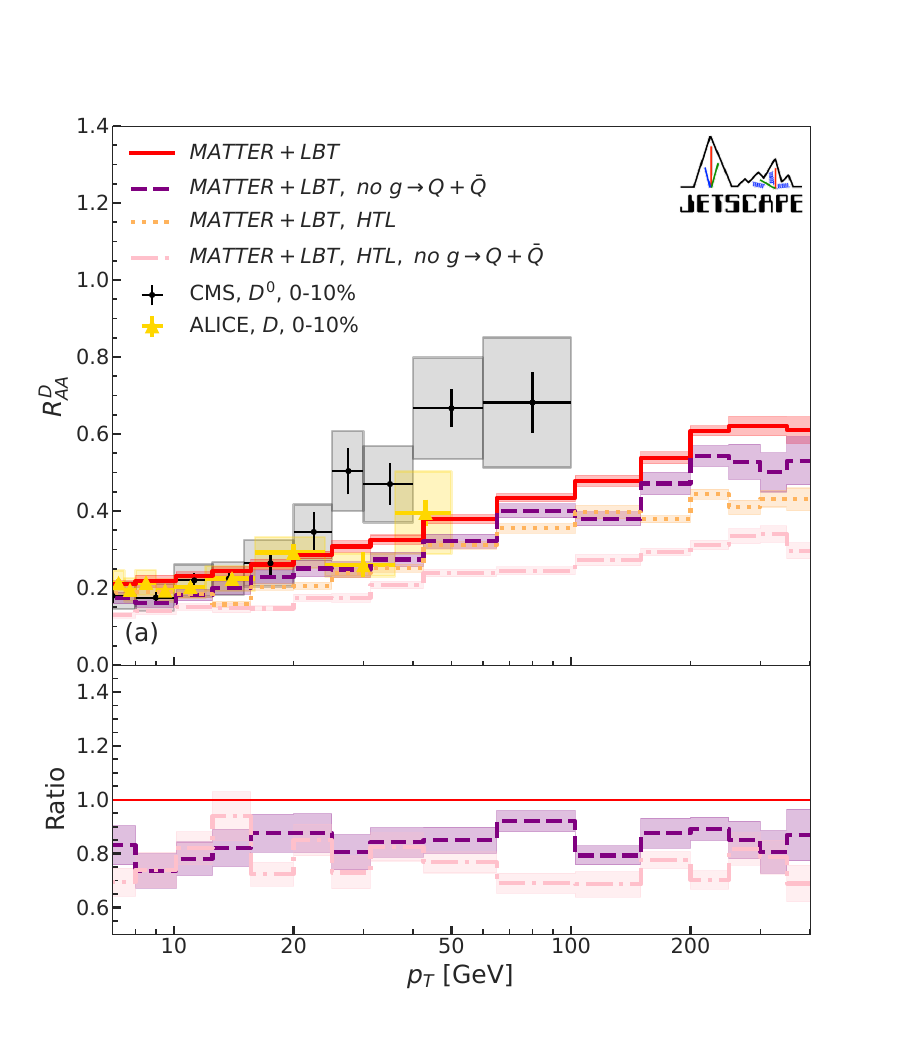} & \includegraphics[width=0.495\textwidth]{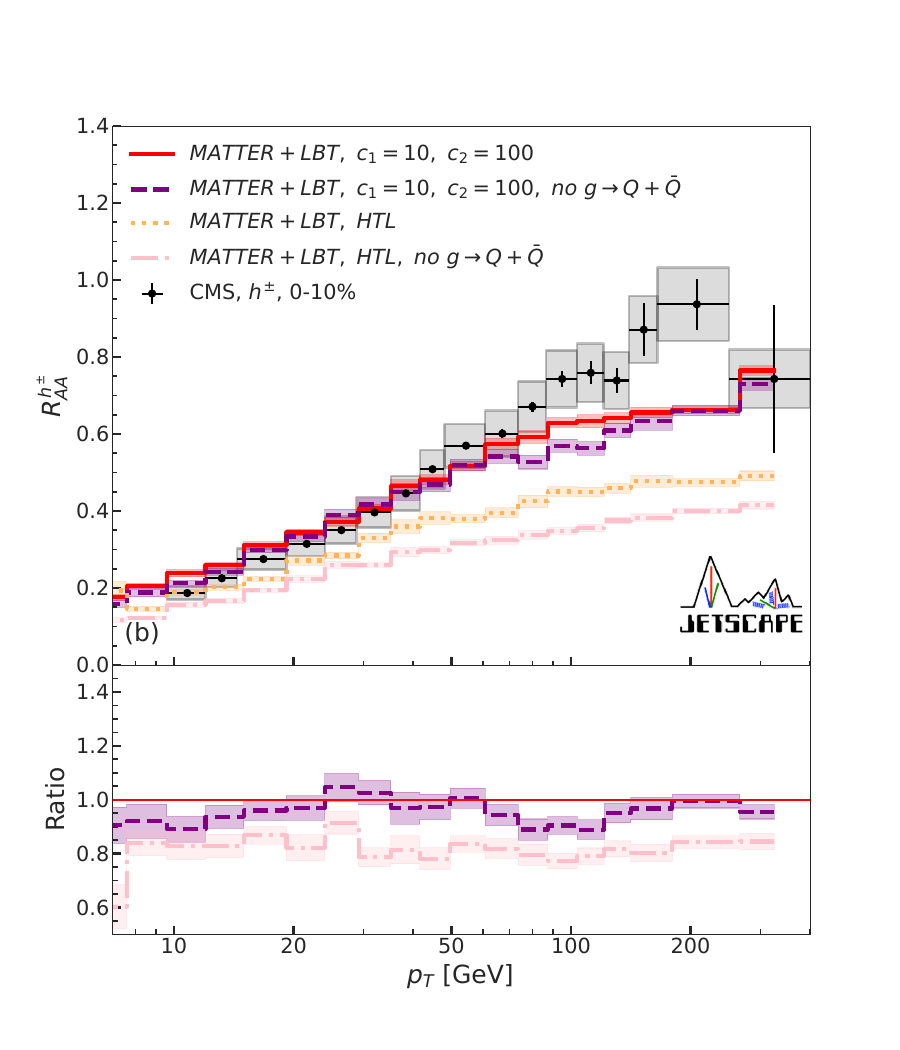}
\end{tabular}
\end{center}
\caption{Nuclear modification factor for D-mesons (a) and charged hadrons (b) in $\sqrt{s_{NN}}=5.02$ TeV Pb-Pb collisions at the LHC at 0-10\% centrality. $c_1=10, \ c_2=100$ parameters values are employed in Eq.~(\ref{eq:qhat_t}). Ignoring the $g\rightarrow Q+\bar{Q}$ process in MATTER impacts the D meson $R_{AA}$, while it has a smaller effect on the charged hadron $R_{AA}$. The dashed line in the ratio subplots divides MATTER+LBT no $g\to Q+\bar{Q}$ to MATTER+LBT, while the dotted-dashed line divides MATTER+LBT HTL no $g\to Q+\bar{Q}$ to MATTER+LBT HTL.}
\label{fig:MATTER_LBT_comp_gQQ}
\end{figure}

Figure \ref{fig:MAT_vs_LBT_all} shows the contribution for MATTER and LBT, as well as the combined result, compared to D-meson and charged hadrons $R_{AA}$, respectively. While it may be possible to change $\hat{q}$ in the LBT-only portion of the simulation to obtain a good description of D-meson $R_{AA}$, such a change would not allow for simultaneous description of both D-meson and charged hadron $R_{AA}$ without a virtuality-dependent $\hat{q}(t)$. The sensitivity to $c_1$ and $c_2$, was well as the importance of having a virtuality-dependent $\hat{q}$ is shown in Fig.~\ref{fig:MATTER_LBT_comp_qhat}, where a similar sensitivity to $\hat{q}(t)$ is observed between light and heavy flavor. Beyond the realization that a non-trivial $\hat{q}(t)$ is required, further constraining $(c_1,c_2)$ is difficult and will be explored via Bayesian analysis \cite{Chen:2023HP}.
Figure~\ref{fig:MATTER_LBT_comp_gQQ} shows the importance of in-medium $g\to Q+\bar{Q}$ to both D-meson and charged hadron $R_{AA}$. According to the ratio plot in Fig.~\ref{fig:MATTER_LBT_comp_gQQ}, the D-meson $R_{AA}$ is significantly affected by $g\to Q+\bar{Q}$, on the order of 20\%. Therefore, any future phenomenological studies of heavy flavor energy loss cannot neglect dynamical $g\to Q+\bar{Q}$ production. The next step for this calculation is to compute the medium-modification to $P_{Q \leftarrow g}$ splitting function, as well as devise a $\hat{q}$ that has both a $t$- and $M$-scale dependence, i.e. $\hat{q}(t,M)$. The formalism to calculate the medium-correction to the $P_{g \leftarrow Q}$ splitting function is already established using Soft Collinear Effective Theory (SCET) \cite{Abir:2015hta}; that approach needs to be adapted for $P_{Q \leftarrow g}$. Furthermore, to obtain $\hat{q}(M,t)$ a modification of the work in Ref.~\cite{Kumar:2019uvu} would be needed. 

%
   
\section{Conclusion}
A realistic calculation of heavy quark interaction within the QGP has been presented, containing both primordial heavy quarks as well as those generated during the shower of a highly virtual parton inside hot and dense nuclear matter. The latter can only be fully described using a multiscale numerical simulation, such as MATTER+LBT. The dynamical production of heavy flavor during the showering of hard partons, i.e. via the $g\to Q+\bar{Q}$ process, contributes significantly (at the 20\% level) to the overall multiplicity of heavy quarks, and thus cannot be ignored. Future directions entail improving the medium-induced $P_{Q \leftarrow g}$ splitting function, and the multiscale dependence of the transport coefficient $\hat{q}(t,M)$, using SCET formalism developed in Refs.~\cite{Abir:2015hta,Kumar:2019uvu} to include both the virtuality and heavy flavor mass scales. 

The calculation presented herein also shows that $\hat{q}(t)$ is needed to describe both light flavor $R_{AA}$ as well as D-meson $R_{AA}$. However to fully constrain the virtuality-dependence of $\hat{q}$, the improvements mentioned above --- i.e. to the $g\to Q+\bar{Q}$ splitting function and $\hat{q}(t,M)$ --- will be needed. Once at hand, they can be applied within a systematic Bayesian approach to constrain $\hat{q}$. The JETSCAPE Collaboration has the necessary tools and know-how to carry out such an endeavor in the future. 

%
\bibliographystyle{h-physrev3.bst}
\bibliography{references}

\end{document}